\documentclass[preprint]{aastex61}
\usepackage{epstopdf}
\usepackage[toc]{appendix}
\usepackage{color}
\usepackage{lineno}
\usepackage{amsmath}
\received{\today}
\submitjournal{ApJ}

\shorttitle{Intense whistler waves in the foreshock transients}
\shortauthors{Shi et al.}

\begin{document}

\title{Intense whistler-mode waves at foreshock transients: characteristics and regimes of wave-particle resonant interaction}  

\correspondingauthor{Xiaofei Shi}
\email{sxf1698@g.ucla.edu}

\author{Xiaofei Shi}
\affiliation{Department of Earth, Planetary, and Space Sciences and Institute of Geophysics and Planetary Physics, \\University of California, Los Angeles, CA, USA}

\author{Terry Liu}
\affiliation{Department of Earth, Planetary, and Space Sciences and Institute of Geophysics and Planetary Physics, \\University of California, Los Angeles, CA, USA}

\author{Anton Artemyev}
\affiliation{Department of Earth, Planetary, and Space Sciences and Institute of Geophysics and Planetary Physics, \\University of California, Los Angeles, CA, USA}
\affiliation{Space Research Institute of the Russian Academy of Sciences, Moscow, 117997, Russia}

\author{Vassilis Angelopoulos}
\affiliation{Department of Earth, Planetary, and Space Sciences and Institute of Geophysics and Planetary Physics, \\University of California, Los Angeles, CA, USA}

\author{Xiao-Jia Zhang}
\affiliation{Department of Earth, Planetary, and Space Sciences and Institute of Geophysics and Planetary Physics, \\University of California, Los Angeles, CA, USA}
\affiliation{Department of Physics, University of Texas at Dallas, Richardson, TX, USA}

\author{Drew L. Turner}
\affiliation{The Johns Hopkins University Applied Physics Laboratory, Laurel, MD, USA}


\begin{abstract}
Thermalization and heating of plasma flows at shocks result in unstable charged-particle distributions which generate a wide range of electromagnetic waves. These waves, in turn, can further accelerate and scatter energetic particles. Thus, the properties of the waves and their implication for wave-particle interactions are critically important for modeling energetic particle dynamics in shock environments. Whistler-mode waves, excited by the electron heat flux or a temperature anisotropy, arise naturally near shocks and foreshock transients. As a result, they can often interact with supra-thermal electrons. The low background magnetic field typical at the core of such transients and the large wave amplitudes may cause such interactions to enter the nonlinear regime. In this study, we present a statistical characterization of whistler-mode waves at foreshock transients around Earth’s bow shock, as they are observed under a wide range of upstream conditions. We find that a significant portion of them are sufficiently intense and coherent (narrowband) to warrant nonlinear treatment. Copious observations of background magnetic field gradients and intense whistler wave amplitudes suggest that phase trapping, a very effective mechanism for electron acceleration in inhomogeneous plasmas, may be the cause. We discuss the implications of our findings for electron acceleration in planetary and astrophysical shock environments.
\end{abstract}


\section{Introduction}
Collisionless shocks are ubiquitous throughout the universe. The plasma reflected by a collisionless shock can stream far away from it along the upstream magnetic field lines. When the field lines are quasi-parallel to the shock normal, a foreshock can form \citep[e.g.,][]{Treumann09}. This is a highly dynamic region just upstream of the shock where reflected particles excite many types of waves and transient structures. In particular observations at Earth's foreshock \citep{Eastwood2005,Gosling1982} have revealed that the foreshock ions can interact with both the solar wind plasma and discontinuities transported by the wind. These interactions result in many types of foreshock transients, including: foreshock bubbles (FBs)\citep{Omidi10,Omidi2020,Turner2020J}, hot flow anomalies (HFAs)\citep{Schwartz18,Lin97:hfa,Lin02:hfa,Omidi&Sibeck07}, and foreshock cavities \citep{Turner13:foreshock,Liu15:foreshock,Schwartz1985,HZhang2022}. 

Shock acceleration is one of the main sources of energetic electrons in astrophysical systems but in order to operate efficiently it requires seed electrons with energies above the injection level (having Larmor radii larger than the shock transition width). However, the pre-acceleration mechanism of electrons to seed-electron energies is still an open question \citep[e.g.,][]{Treumann09}. Previous observations suggest that foreshock transients may be important for pre-accelerating electrons, and can therefore contribute significantly to shock acceleration \citep{wilson2016_foreshock}. They can do so, e.g., by capturing ambient foreshock electrons and further energizing them through betatron acceleration \citep{Liu19:foreshock}. As a foreshock transient boundary convects earthward (towards the bow shock), particles inside the core could gain additional energy through Fermi acceleration \citep{Liu17:foreshock&electrons,Turner2018, Omidi2021}. In fact, a recent statistical study showed that electrons are almost always accelerated inside the core region of foreshock transients \citep{Liu2017}. For these reasons, it is important to further explore the electron acceleration mechanisms in foreshock transients.  

Around the bow shock and inside foreshock transients there are many field fluctuations and waves that could either directly accelerate electrons or modulate other acceleration processes (e.g., Fermi and betatron acceleration) \citep[e.g.,][]{Oka19, Lichko&Egedal20}.
One of the most effective wave modes for electron scattering and acceleration is electromagnetic whistler-mode waves \citep{Gary:book05,bookGurnett&Bhattacharjee05}. These can be generated by the heat flux anisotropy \citep{Gary&Feldman77:heatflux} or temperature anisotropy of electron distributions \citep{Sagdeev&Shafranov61, Kennel66}. There is good evidence that both of these types of anisotropy can arise when solar wind electrons interact with the bow shock and foreshock transients \citep[e.g.,][and references therein]{Vasko20:pop,Page&Vasko21}. The importance of whistler-mode waves for electron scattering and acceleration at the bow shock has been extensively investigated and discussed \citep{Hull12,Oka17,Oka19,Amano20}. The role of these waves on electron energization around foreshock transients, however, remains to be fully understood. Although previous case studies have shown that whistler waves exist in foreshock transient environments \citep{Wilson13} and may effectively scatter and accelerate electrons  \citep{Shi20:foreshock_whistlers,Artemyev22:jgr:foreshock}, their occurrence rate, spatial distribution, and wave properties (propagation, polarization, intensity) have not yet been studied comprehensively, for a range of events and plasma conditions. It is clear that a statistical study aimed at determining the properties and the potential contribution of whistler-mode waves for electron acceleration on foreshock environments is timely and important.

The desired statistical study of wave properties would ideally distinguish the dominant regime of wave-particle resonant interactions. Low amplitude, broad-band waves scatter electrons in the diffusive regime of resonant interactions \citep{Kennel69, Lyons72, Veltri&Zimbardo93,Amano20}, commonly observed in the solar wind \citep{Tong19:ApJ,Verscharen22}. Sufficiently intense, narrow band (coherent) waves may resonate with electrons in the nonlinear regime of resonant interactions \citep{Shklyar09:review, Albert13:AGU, Artemyev18:cnsns}. Nonlinear resonant interactions include the phase trapping mechanism \citep{ONeil65,Nunn71}, which is quite effective for electron acceleration in Earth's outer radiation belt  \citep{Chernikov92,Ucer01,Kuramitsu05}. Once the regime of nonlinear resonant interactions with electrons can be statistically established for the whistler-mode waves of interest at foreshock transients, then their investigation can proceed using the formalism developed (and statistical studies conducted) in a similarly inhomogeneous magnetic field environment, Earth's inner magnetosphere \citep{Karpman74,Inan&Bell77,Solovev&Shkliar86,Albert93}.

\begin{figure}
\centering
\includegraphics[width=0.6\textwidth]{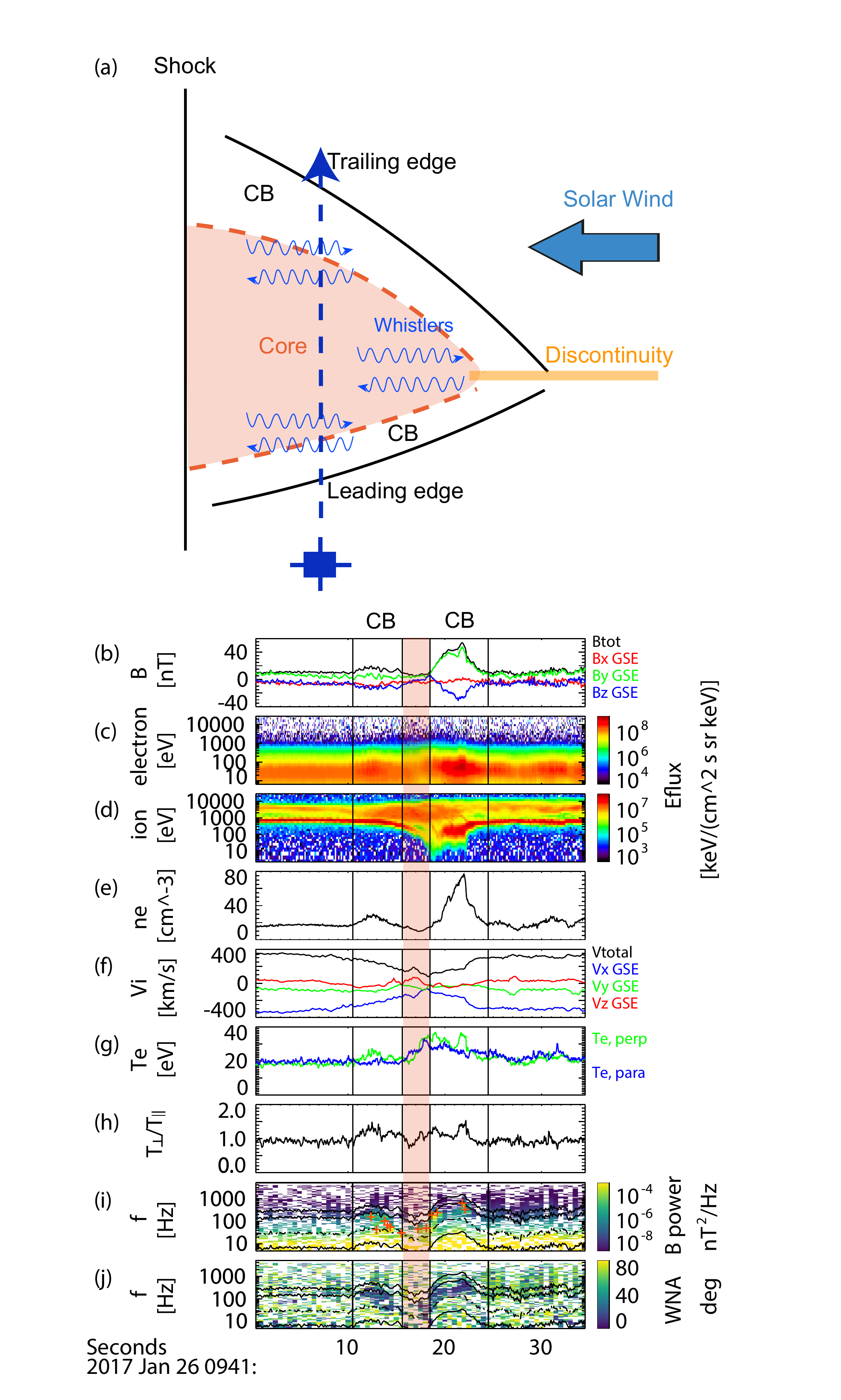}
\caption{\label{fig1} (a) Sketch of an HFA cross-section in the HFA reference frame, at an instant in time as it moves along the shock, past a spacecraft (down). The HFA's hot plasma core is flanked by compressional boundaries (CB). Whistler waves are observed around the edge of the core. The blue dashed line shows the spacecraft trajectory (up) relative to the HFA in this frame.  (b) Magnetic field, (c,d) Energy flux of electrons and ions, (e) Electron density, (f) Plasma velocity, (g) Electron perpendicular temperature and parallel temperature, (h) Electron temperature anisotropy ($T_\perp/T_\parallel$), (i) Magnetic field power spectral density (red crosses depict mean frequency determined as discussed in the text), and (j) Wave normal angle. The core of the HFA is marked by the orange shaded region.}
\end{figure}

We focus on whistler-mode waves observed around HFAs and FBs, the types of foreshock transients with the most significant plasma and field fluctuations. Whistler-mode waves occur regularly within and around them. Such transients occur at least ten times per day, particularly during conditions of the above-average solar wind speed \citep{Lu2022,Chu2017}. Both transient types have a hot, tenuous core associated with strong plasma deflection. HFAs are surrounded by compressional boundaries on either side, while FBs have an upstream compressional boundary bounded by their own mini-shock. They have a scale of one to several Earth radii ($R_E$). Figure \ref{fig1} (a) shows a sketch of an HFA that forms in response to an approaching solar wind discontinuity. The hot plasma generated in the core expands and the large-gyroradius hot foreshock ions at its edge form the core compressional boundaries at its two sides. Whistler-mode waves are observed in the core and compressional boundaries.  The HFA moves (slides) anti-sunward along the bow shock (downward in the figure); the dashed blue line shows the spacecraft trajectory relative to the HFA.

Because foreshock transients occur frequently, at increasing occurrence rates with solar wind Mach number, and with intense whistler-mode waves have been observed within them, such structures are an important ingredient of shock environments at Earth and by inference in all astrophysical systems which are expected to harbor high-Mach-number, quasi-parallel shocks. The effect of foreshock transients on electron pre-acceleration, particularly the large amplitude whistlers within, has yet to be explored. This study will statistically assess the potential of whistler waves for electron acceleration and scattering in foreshock transients. Comprehensive (multi-instrument, multi-point), in-situ measurements in Earth’s bow-shock enable us to examine these waves in great depth and draw conclusions for their astrophysical counterparts of similar upstream Mach number, plasma beta, and normalized spatial extent conditions.

\section{Data and methodology}
\label{sec:data}

We used data from the Magnetospheric Multiscale (MMS) mission which consists of four identical satellites in a tetrahedral configuration \citep{Burch16}. The low frequency magnetic field is measured by the fluxgate magnetometer (FGM) \citep{Russell16:mms} and the high frequency by the search coil magnetometer (SCM)\citep{LeContel16} at a rate of 128 S/s and 8192 S/s, respectively, while in burst mode (as is the case here). The fast plasma investigation (FPI)\citep{Pollock16:mms} instrument provides ion and electron measurements at a resolution of 150ms and 30ms, respectively (also in burst mode).

We identified 208 HFA and FB events in the MMS data collected between 2017 to 2022. They were selected by the following criteria: (1) they have a hot core with one or two compressional boundaries; (2) inside the core, density, velocity, and field strength are reduced, but temperature is increased; (3) compressional boundaries are accompanied by a sharp increase in magnetic field strength and density. Events that are overly complex and lack a clear transient structure are excluded (e.g., multiple HFAs, whose boundaries are unclear). Our criteria, therefore, favor isolated events. Figure \ref{fig1}(b-f) represents an observation of a typical HFA with two compressional boundaries. Quasi-parallel propagating whistler waves, evidenced by an increase in wave power  are observed within the compressional boundaries and on the edge of the core (Figures \ref{fig1}i-j). The observed whistler waves coincide with increases in the electron perpendicular temperature anisotropy (Figures \ref{fig1} (g) and (h)). 

We use the power spectral density (PSD) to determine the mean frequency ($\langle f \rangle$) and the frequency width ($\Delta f$) of whistler waves:

\begin{equation}
\langle f \rangle=\frac{\int_{f_{low}}^{f_{ce}}PSD_wfdf}{\int_{f_{low}}^{f_{ce}}PSD_wdf}
\end{equation}

\begin{equation}
(\Delta f)^2=\frac{\int_{f_{low}}^{f_{ce}}PSD_w(f-\langle f \rangle)^2df}{\int_{f_{low}}^{f_{ce}}PSD_wdf}
\end{equation}
where $PSD_w(f)=PSD(f)-PSD_b(f)$, and $f_{low}=max\{4Hz, f_{lh}\}$. $PSD$ is averaged over every 0.5s and $PSD_b$ is the background power spectrum (partly due to instrument noise), subtracted here in order to detrend the spectrograms and better reveal the waves. This background spectrum was obtained at each frequency by averaging all times when the $PSD$ fell to $<30\%$ of its average in each event. Using the wave frequency $\langle f \rangle$ and the frequency width $\Delta f$, we compute the average wave amplitude ($\langle B_w\rangle$) and maximum wave amplitude ($B_w$) from band-pass filter data in each 0.5s wave interval. 

Using timing and MVA (minimum variance analysis, see \citet{Sonnerup68,bookISSI:Sonnerup}) methods to determine the magnitude and direction of the wave vector ($\vec{k}$), we then calculated the wave normal angle and the wave frequency in the plasma frame by correcting for its Doppler-shift relative to the spacecraft frame, where we measured it. All wave properties have been averaged every 0.5s. MMS provides four-point observations, from close separations. As the plasma frequency ($f_{pe}$) in foreshock transients is usually 100-200 times larger than the electron gyrofrequency ($f_{ce}$), the wave vector ($k=\frac{2\pi f_{pe}}{c}\sqrt{f/(f_{ce}\cos{\theta_{kb}}-f)}\sim f_{pe}/c$) is large and the wavelength (about 10s of km) is $\lesssim$ to the average separation between the MMS satellites. Therefore, the timing method can be applied to directly obtain the value and direction of $\vec{k}$ \citep{Paschmann,Turner2017wave}. For a coherent wave signal, $\vec{k}$ obeys the following linear equations:

\begin{equation}
\begin{pmatrix}
R_{12x} & R_{12y} & R_{12z} \\
R_{13x} & R_{13y} & R_{13z} \\
R_{14x} & R_{14y} & R_{14z} 
\end{pmatrix}
\begin{pmatrix}
k_x \\
k_y \\
k_z 
\end{pmatrix}
=
\begin{pmatrix}
\Delta\phi_{12} \\
\Delta\phi_{13} \\
\Delta\phi_{14} 
\end{pmatrix}
\end{equation}
where $R$ is the separation between two satellites;  $\Delta\phi=2\pi\delta t/T$ is the phase difference between two satellites, where $\delta t$ is the lag time corresponding to the peak cross-correlation and $T$ is the observed period of the wave. The peak cross-correlations between the wave fields observed by four satellites allow us to assess the accuracy of the results. Only events with all inter-satellite peak cross-correlations larger than $0.8$ were kept. For those events we then calculated the mean wave frequency in the plasma frame: $2\pi f=2\pi\langle f\rangle-\vec{k}\cdot\vec{v_p}$, where $\vec{v_p}$ is the plasma (ion) velocity.

While the timing method, used above, can directly measure the magnitude and absolute direction of $\vec{k}$ but requires coherent four-point measurements of the wave fields. For this reason, we also used the one-point MVA technique to estimate the orientation of $\vec{k}$ when the timing method is not applicable. The MVA method calculates the principal variance directions and their associated eigenvalues \citep{Paschmann}. The direction of $\vec{k}$ is the minimum variance direction. To ensure that the minimum variance direction is well determined and the waves are circularly polarized, we only kept the points with $\lambda_{int}/\lambda_{min}>10$ (the ratio of intermediate to minimum eigenvalues) and $\lambda_{max}/\lambda_{int}<3$ (the ratio of maximum to intermediate eigenvalues). To compute $|\vec{k}|$ for the waves of interest we applied the MVA method to band-pass filtered data around the mean frequency (in the range $[\langle f \rangle-\Delta f, \langle f \rangle+\Delta f]$) \citep{Wilson13:waves}. The $180^\circ$ ambiguity in the $\vec{k}$ direction in the MVA method can be eliminated by assuming that the Poynting vector $\vec{S}$ and $\vec{k}$ are roughly in the same direction \citep{Verkhoglyadova10,Verkhoglyadova13} -- we computed  $\vec{S}$ using both electric and magnetic field data, in Appendix A from \cite{Wilson13:waves}, for all our events and determined the sign of wave propagation for MVA-computed $\vec{k}$ values. We then used this $\vec{k}$ to compute the Doppler-shifted wave frequency in the plasma frame from the MVA method.

\section{Statistical results}

\begin{figure}
\centering
\includegraphics[width=1\textwidth]{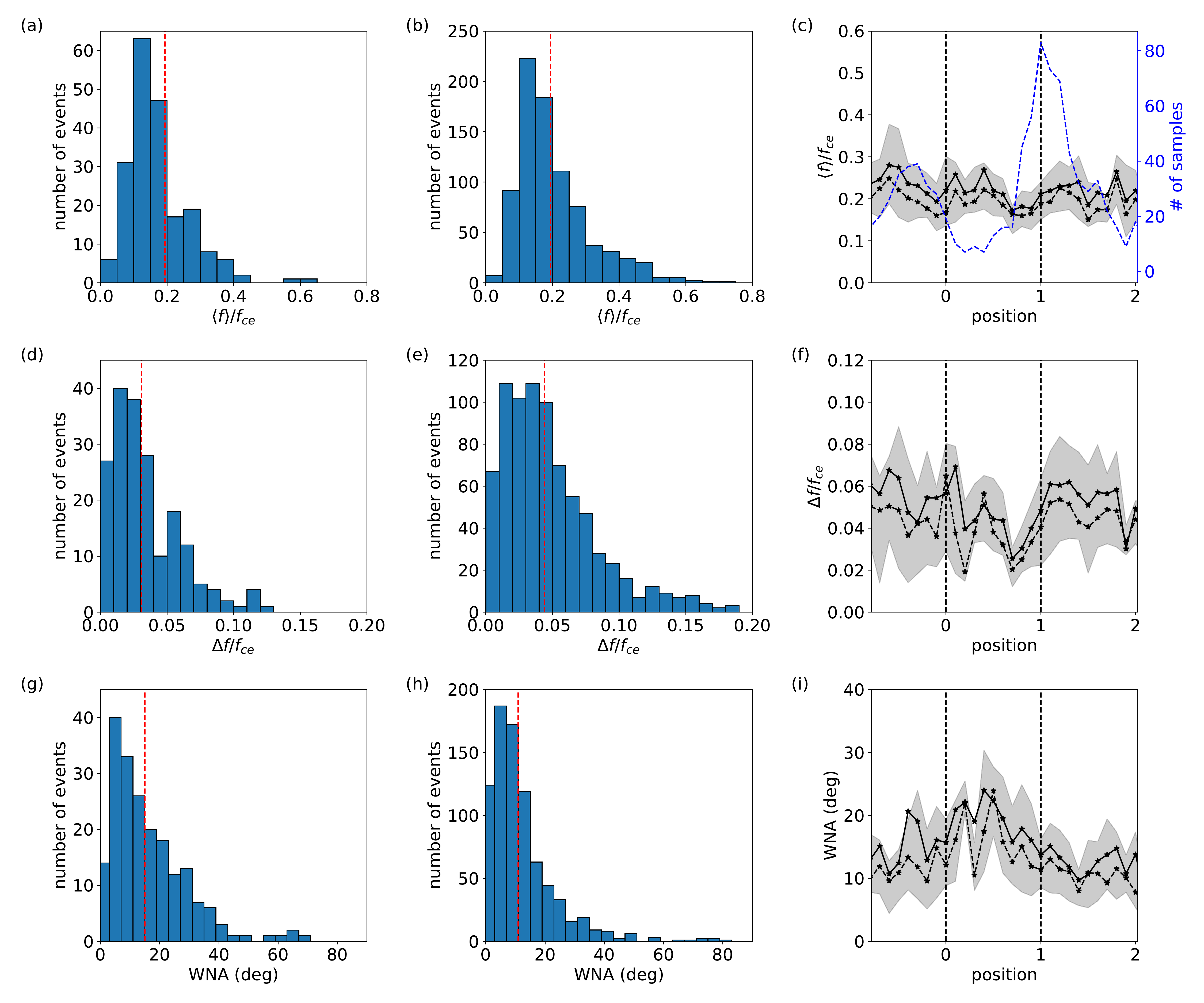}
\caption{\label{fig2} The number histograms and spatial distribution of: (a-c) wave frequency; (d-f) wave frequency width; and (g-i) wave normal angle. The left and middle columns show histograms in the core and in the compressional boundaries, respectively. Dashed red lines are medians. The right column shows average and median values, in solid black lines and dashed black lines, respectively; the lower and upper bound of the shaded region represents the $25^{th}$ and $75^{th}$ percentile of the data, respectively; the dashed blue line in Panel (c) is the number of whistler waves observed at different locations.}
\end{figure}

\begin{figure}
\centering
\includegraphics[width=1\textwidth]{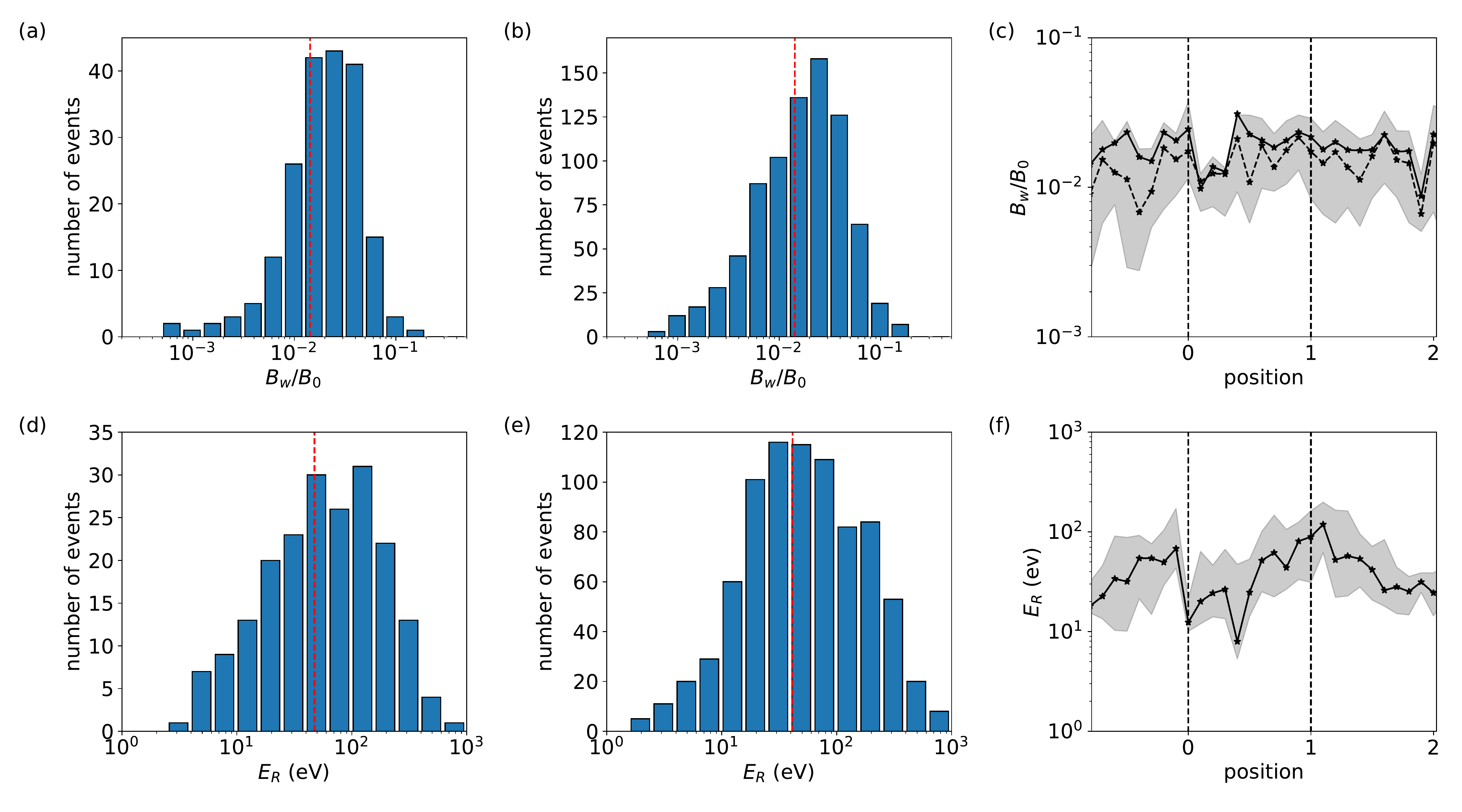}
\caption{\label{fig3} Number histograms and spatial distribution of (a-c) maximum wave amplitude, and (g-i) minimum resonance energy for the mean wave frequency. Dashed red lines show the median value.}
\end{figure}

 Whistler waves were observed in 85$\%$ of all foreshock transient events in our database. The wave spatial distribution within the foreshock transients (in the core or the compressional boundary) is important because it highlights where the waves are preferentially generated and where they may interact with electrons. To reveal the spatial distribution of whistler waves, we normalized the time interval of the core region to $[0,1]$, based on crossing times specific to each event. The leading and tailing boundaries were located at normalized times less than 0 and larger than 1, respectively. As mentioned previously, the compressional boundary forms along the edge of the foreshock and is characterized by the enhanced magnetic field and plasma density. We defined the edge of the boundary by where the magnetic field magnitude equals the background value, where the background field was calculated by averaging the magnetic field strength in the relatively quiescent region upstream of each foreshock transient. (For example, in Figure \ref{fig1} (b), the edges of the compressional boundary are shown by the vertical lines.) Note that FBs usually only have one trailing boundary. Therefore, the normalized FB events start at position 0.  Figures \ref{fig2} and \ref{fig3} depict the superposed epoch analysis of whistler wave properties and their spatial distribution in foreshock transients versus the normalized time (to be interpreted as the spatial location within the core, or relative distance from the core boundaries). 

Figures \ref{fig2}(a,b) show the number histograms of the normalized mean frequency ($\langle f\rangle /f_{ce}$) in the core and in the compressional boundary regions of our events, respectively. The median frequency in both regions is around $0.2f_{ce}$. The spatial distribution (versus normalized time) of $\langle f \rangle/f_{ce}$ is shown in Figure \ref{fig2}(c). The solid and dashed black lines are the mean and median values of $\langle f \rangle/f_{ce}$; they are $\sim 0.2f_{ce}$ at all locations, despite the sharp change of the background magnetic field strength and density at the compressional boundary. This indicates a local generation mechanism for most of the observed waves. There is also a local peak around $\langle f \rangle/f_{ce}\sim 0.3$ observed in the core (see Figure \ref{fig2}(a)) but not in the boundary. This implies that some fraction of the waves observed in the core, especially with $\langle f \rangle/f_{ce}\in[0.3,0.5]$, may be generated within the boundaries and then propagate to the core region. Because the core region is characterized by smaller background magnetic fields, when whistler waves propagate into it, their relative frequency ($\langle f \rangle/f_{ce}$) increases. We estimated that these waves with $\langle f \rangle/f_{ce}\in[0.3,0.5]$ represent less than $15$\% of waves observed within the core region.  

The dashed blue line in Figure \ref{fig2}(c) shows the number of events with whistler waves as a function of position. It is equivalent to the spatial statistical distribution of the waves as a function of location within the foreshock transient. It shows that whistler-mode waves are most probable near the edge of the core where the magnetic field gradient is large. More events are located near position=1 versus position=0 because we included FBs, which usually only have a single compressional boundary, on the trailing side. 

Figures \ref{fig2} (d-f) show the number histograms and spatial distribution of the normalized wave frequency width $(\Delta f/f_{ce})$ in a similar manner as the frequency panels, Panels (a-c) above. The median value of $\Delta f/f_{ce}$ is $\sim 0.03$ in the core and $\sim 0.04$ in the compressional boundaries. The frequency width does not depend on the positions. The observed waves are quite narrow-banded, with $\Delta f/f\sim 1/5$, which suggests a narrow resonant energy range of electrons responsible for wave generation, i.e. the anisotropic electron population is bounded below and above in energy by isotropic cold and hot electrons, respectively \citep[e.g.,][]{Fu14:radiation_belts,Page&Vasko21,Frantsuzov22}.

The number histograms and the spatial distribution of wave normal angles are shown in Figures \ref{fig2}(g-i) in a similar format as the panels above. Most whistler waves are quasi-parallel propagating, and the waves tend to be more oblique in the core region: the medians increase from $12^\circ$ in the boundary to $16^\circ$ in the core (Panels (h) and (g)), and this is also evident in the spatial profile of both the medians and means in Panel (i). This is likely a result of (1) wave propagation to the core region from the boundaries (whistler wave propagation in inhomogeneous magnetic field and plasma results in a wave normal angle increase, see \citet{Shklyar04,Chen13,Gu21}) or (2) oblique wave generation within the core by either cyclotron or Landau resonance thanks to the suppression of Landau damping by the large parallel temperature often observed in that region (see discussion of such oblique wave generation by, e.g., \citet{Li16}).

Number histograms and spatial distribution of the normalized maximum wave amplitude ($B_w/B_0$) in our events are shown in Figures \ref{fig3}(a-c), in a format similar to \ref{fig2}. Here, $B_0$ is the background magnetic field strength -- typically $\sim 5-10s$ nT in HFAs and FBs. The median value for $B_w/B_0$ in both core and compressional boundaries is larger than 0.01. Thus, the maximum wave amplitude can reach $10s-100s$ of pT. The mean wave amplitude (not shown here) is about 3 times smaller than the maximum values in statistics. Figures \ref{fig3} (d-f) show the distribution of electron minimum resonance energies ($E_R$) for the mean wave frequencies (we used equation (B1) from \cite{Wilson13:waves} to calculate the resonance energy). Whistler waves are mainly resonant with electrons which have $10s-100s$ eV parallel energy. This is the hot solar wind electron population, having energies larger than the typical solar wind temperature. 

The electron resonant interaction with whistler-mode waves is controlled by the wave amplitude, wave spectral width $\Delta f$, and background field inhomogeneity $\partial B_0/\partial s$ ($s$ is the distance along the field from the equator and is also used to mark the location of the interaction along the field-line direction). If the wave amplitude is low the waves cannot alter the electron orbit significantly during a single resonant interaction and the interaction remains first-order (linear) and can be described well by zero-order orbit perturbation theory \citep[quasi-linear diffusion is a particular version of this theory see][]{Kennel&Engelmann66}. If the wave spectral width is too broad relative to the resonance width, nonlinearity from that resonance can also be evaded. These two conditions can be expressed as two criteria for nonlinear interaction. The first assesses the nonlinearity for a pure mode using the inhomogeneity parameter, S, a function of the normalized wave amplitude $B_w/B_0$ and $\partial B_0/\partial s$. The second criterion on the spectral width $\Delta f$ addresses the spectral purity (the monochromatic nature) of the wave.

For highly coherent (approximately monochromatic, or pure mode) waves in an inhomogeneous magnetic field, the nonlinearity criterion for $S\propto  (\partial B_0/\partial s)/(B_w/B_0)$ \citep{Omura08} is:

\begin{equation}
S=\frac{1}{2}\frac{N^2}{N^2-1}\left(\frac{kv_{\perp}}{f_{ce}}-\left(3-\frac{1}{N^2}\right)\frac{v_R}{v_{\perp}}\right)\cdot\left(\frac{1}{kB_0}\frac{\partial B_0}{\partial s}\right)\cdot\frac{B_0}{B_w},
\end{equation}
where $k$ the value of wave vector, $N=kc/\omega$ the wave refractive index, $v_{\perp}=\sqrt{v^2-v_R^2}$ the electron transverse velocity, $v_R=2\pi(f-f_{ce})/k$ the resonant velocity (where $k$ is in rads/km), and $\partial/\partial s$ the gradient along magnetic field lines. The inhomogeneity of the background magnetic field ($\partial B/\partial s$) is computed using the linear estimation of the gradient method from four-satellite observations (see Chapter 14 of \cite{Paschmann&Schwartz00}). When $|S|<1$ the wave is sufficiently strong to locally overcome the mirror force $\sim \partial B_0/\partial s$ and alter the electron trajectory significantly. This is the regime of nonlinear resonant interactions. $S$ depends on the electron energy and pitch angle (linked by the resonance condition). For $\partial B_0/\partial s \ne 0$, S is finite and can be evaluated in two limits: (1) in the local limit, $S_{local}$ evaluated explicitly shows if electrons with a given energy and pitch angle will interact with waves nonlinearly, and (2) in the global limit, $S_{global}$ can be evaluated by projecting the electron pitch angle ($\alpha$) from the location $s$ of the wave measurements to the location where $B_0$ reaches its minima. In the global limit, $S_{global}$ captures how often along their zero-order, adiabatic trajectory ($\sin^2\alpha/B_0=const$) electrons will interact with the waves nonlinearly. 

For each wave event in our database, we obtained the 0.5-second averaged value of the wave properties ($\vec{k}$, $\langle f \rangle$, $B_w$), electron cyclotron frequency $f_{ce}$, background field $B_0$, background inhomogeneity $\partial B/\partial s$, and electron parallel resonant velocity $v_R$. To compute $S$, we also need $v_{\perp}$.  Cyclotron resonance happens when the electron parallel velocity is $v_{\parallel}=v_R$. Therefore $v_{\perp}=\sqrt{v^2-v_R^2}=v\sin{\alpha}$ is a function of the (total) resonance energy ($E=1/2m_ev^2$) and $\alpha$. (Note that the total energy and $\alpha$ are connected by $\cos{\alpha}=v_R/v$). We used this to compute S in the local limit, and from it S in the global limit. 

\begin{figure}
\centering
\includegraphics[width=0.8\textwidth]{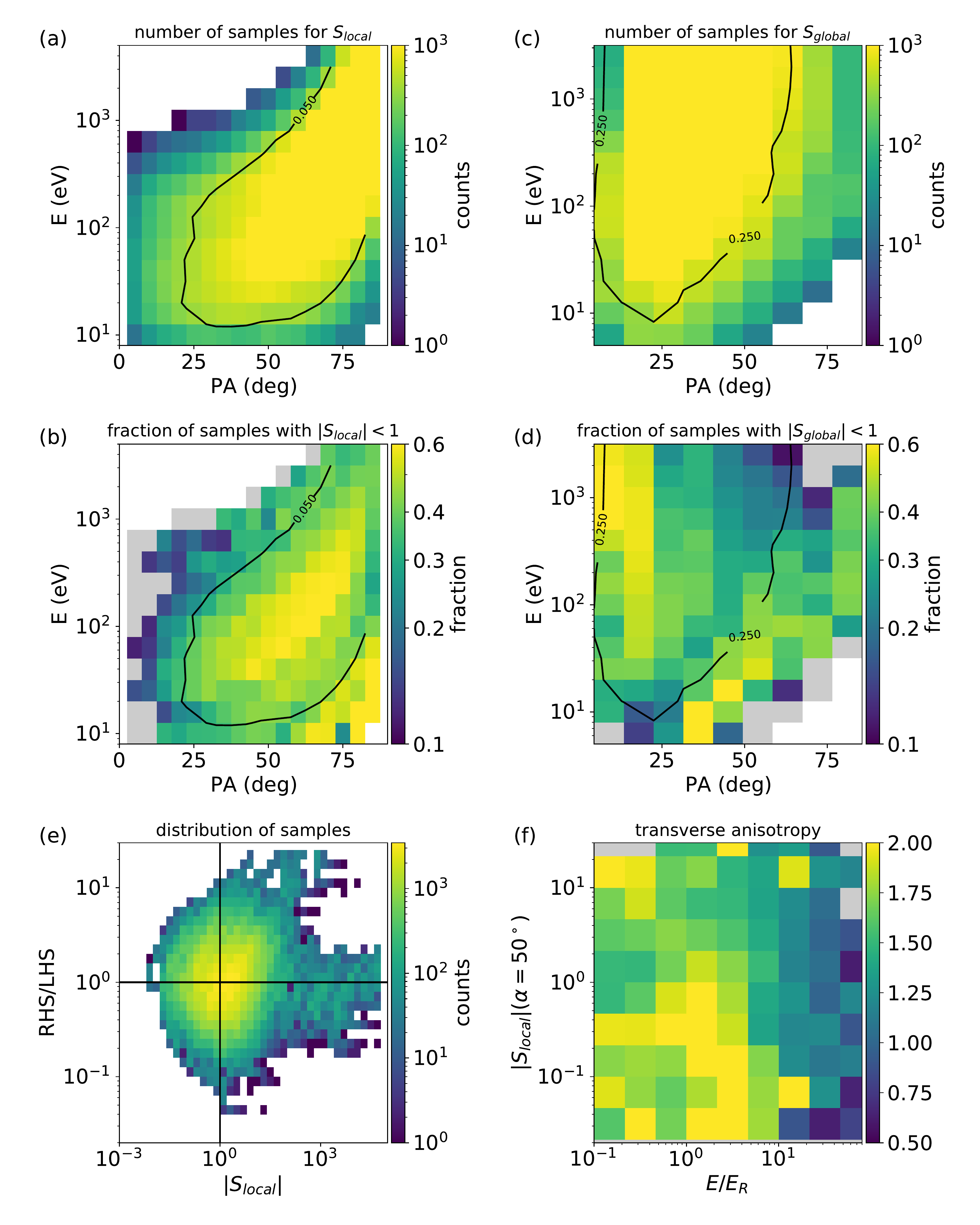}
\caption{\label{fig4} Panel (a, b) shows the distribution of samples for $S_{local}$ and $S+{global}$ calculation, respectively. The black contour shows the 95\% of total samples. Panel (c) shows the sample fraction of whistler waves resonating with electrons in the nonlinear regime ($|S_{local}|<1$). Panel (d) repeats panel (c), but for the measurements with pitch-angle adiabatically projected to the minimum of the background magnetic field from the wave observation location ($S_{global}$). Panel (e) shows the number of wave observations versus $(|S_{local}|,RHS/LHS)$, where RHS and LHS refer to the right-hand side and left-hand side of Eq (5), respectively. $|S_{local}|<1$ and $RHS/LHS>1$ define the region of nonlinear wave-particle interactions (see text for details). Panel (f) shows the distribution of electron flux anisotropy versus $|S_{local}|$ and energy. Energy is normalized to the resonance energy for the observed waves.}
\end{figure}

We computed $S_{local}$ at different energies using the (local) measurements. Combining all the local measurements, we can arrive at the number distribution and the value of $S_{local}$ versus energy and pitch angle in our dataset. Figure \ref{fig4}(a) shows the distribution of the number of measurements used for this computation. The black contour shows the number of samples per bin below which lie just $5$\% of all observations, i.e., the number density of measurements outside this contour is rather low (interpreted as being insignificant) by comparison to the rest of the parameter space inside that contour. Figure \ref{fig4}(c) shows the fraction of measurements with $|S_{local}|<1$; it represents the probability distribution for electrons to interact with waves nonlinearly in the local limit. The $5$\% contour of the number of measurements is transferred from the panel above here and demarcates the region within which the probability distribution is trustworthy. We see that in the region of $\alpha>45^\circ$ and energy within $[300,1000]$ eV the fractions are high, $>30\%$ and can get up to $\sim60\%$. In this region of (E,$\alpha$) space with a sufficient number of measurements, there is a large enough probability for $|S_{local}|$ to be $<1$ (for the observed waves to be sufficiently intense) such that the waves interact with electrons nonlinearly.  

The global limit is obtained under the assumption that electrons are bouncing within a local magnetic field trap. In each event, we projected the local electron pitch angle to the location where $B_0$ reaches its minimum (around the center of the core region), and then obtained $S_{global}$ by mapping $S_{local}$ to a new pitch angle corresponding to that minimum. Such mapping also removes the direct connection between the electron energy and (mapped) pitch-angle through the resonant condition. The number of measurements in energy versus mapped pitch-angle space is shown in Figure \ref{fig4}(b) in a format similar to that of Figure \ref{fig4}(a), including the contours. The resultant fraction of $|S_{global}|<1$ is shown in Figure \ref{fig4}(d), with contours transferred from the panel above it. It shows that $\alpha<45^\circ$, and $>100$ eV electrons will be resonant with whistle-mode waves nonlinearly (note measurements outside of $75$\% contour are not statistically representative).

Overall, Figures \ref{fig4}(b,d) demonstrate that in the foreshock transients of our database, quite often the whistler waves are strong enough to cause nonlinear resonant interaction with electrons; the background magnetic field inhomogeneity is too weak to suppress this nonlinear behavior. However, the wave spectrum width $\Delta f$ also influences the resonant interactions. The above considerations assume that the waves are sufficiently monochromatic, i.e., $\Delta f$ is small enough. The criterion for small $\Delta f$ can be derived from \citep{Karpman74:ssr,LeQueau&Roux87}:

\begin{equation}
\frac{\Delta f}{\langle f \rangle}<\left(\frac{B_w}{B_0}\frac{vf_{pe}}{cf_{ce}}\right)^{1/2}\left(\frac{f/f_{ce}}{1-f/f_{ce}}\right)^{1/4}
\end{equation} 
where $v$ is electron velocity determined by the resonance condition ($v=v_R/\cos{\alpha}$). If this criterion is not satisfied, the wave spectrum is broadband (random phase approximation is valid) to break nonlinear resonance effects. If the criterion is satisfied, the phase of waves in the wave packet can be coherent and prevent random phase mixing. Similar to the $S_{local}$ treatment, we statistically collected the distribution of $RHS/LHS$  versus energy and pitch angles for each local measurement, where $RHS$ and $LHS$ are the right-hand side and left-hand side terms of Eq.(5). Then we rearranged the data to get the distribution of the wave measurements versus $(|S_{local}|, RHS/LHS)$ (shown in Figure \ref{fig4}(e)). The region $|S_{local}|<1$ and $RHS/LHS>1$ corresponds to that of nonlinear resonant interactions. The requirement of a narrow wave spectrum halves the number of observed waves resonating with electrons nonlinearly, i.e., $\sim 30$\% of observed waves have $|S_{local}|<1$, but only half of these waves have $RHS/LHS>1$. But even $\sim 15$\% of waves populating  $|S_{local}|<1$ and $RHS/LHS>1$ region provides a sufficiently large occurrence rate of nonlinear wave-particle interactions. This occurrence rate is comparable to (or even larger than) that of electron nonlinear resonances with whistler-mode waves in the Earth’s radiation belts \citep{Zhang19:grl} and is much larger than the occurrence rate of electron nonlinear resonances with whistler-mode waves in the solar wind \citep{Tong19:ApJ}.

To identify the possible source of such intense whistler-mode waves resonating with electrons nonlinearly, we examined the electron distribution function (DF) anisotropy by evaluating the transverse-to-parallel phase space density ratio, or transverse anisotropy, $DF_\perp/DF_\parallel$. For each local measurement, we computed this ratio at all different energies. We also computed the measurement's $S_{local}$ value assuming a fixed, representative value for $\alpha= 50^\circ$ (corresponding to a significant fraction of waves with $|S_{local}|<1$, based on Figure \ref{fig4}(c)). Figure \ref{fig4}(f) shows the median of the aforementioned electron transverse anisotropy as a function of normalized energy $E/E_R$ and $|S_{local}|$ value. Near the resonance energy, $E/E_R\in [0.1,10]$, the electron anisotropy for intense waves (those with $|S_{local}|<1$)  maximizes and reaches $\sim 2$. Such a high electron anisotropy should result in large whistler-mode wave growth rates and large wave amplitudes. Although the quantity $DF_\perp/DF_\parallel$ cannot uncover an electron heat flux anisotropy (one of the important free energy sources for whistler-mode waves via the heat flux instability \citep[e.g.,][]{Gary&Feldman77:heatflux,Tong19:ApJL}), a large value of this transverse anisotropy can either result in amplification of waves generated by the heat flux instability or it can reveal the primary source for wave generation \citep{Vasko20:pop}. Thus, Figure \ref{fig4}(f) shows that intense whistler-mode waves driven (or amplified) by a large electron transverse anisotropy (1-2) may resonate with electrons nonlinearly ($|S_{local}|<1$). However, it is noteworthy that some of the large transverse anisotropies observed at high energies ($E/E_R \sim 10$ may also be generated by electron nonlinear resonant acceleration by whistler-mode waves.

With such a significant population of whistler-mode waves resonating nonlinearly with electrons, we anticipate the electron distributions to exhibit signatures of these interactions. Motivated by the discussion above, we separated the linear and nonlinear wave-particle interactions as follows: (1) linear regime with $|S_{local}|>1$ or  $RHS/LHS<1$,  and (2) nonlinear regime with $|S_{local}|<1$ and $RHS/LHS>1$. 

\begin{figure}
\centering
\includegraphics[width=0.8\textwidth]{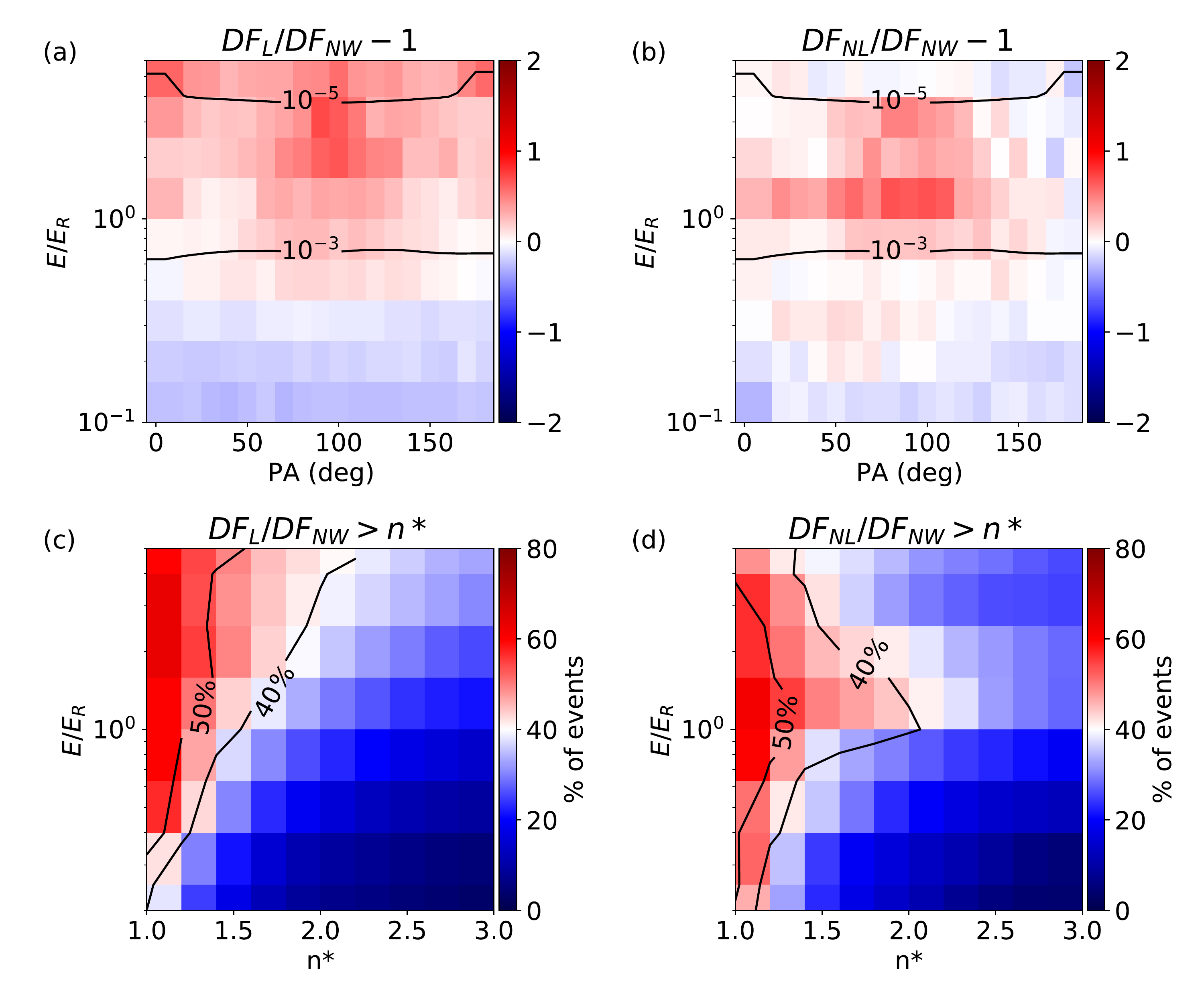}
\caption{\label{fig5} Electron distribution functions (DFs) collected during times when waves were observed, normalized to those collected in the absence of waves. Panels (a) and (b) show electron DFs associated with weak (a) and intense (b) wave observations normalized to the background (measured in absence of waves; subindex $NW$) electron spectra. Panels (c) and (d) show the probability distributions of $DF_L/DF_{NW}>n^*$ and $DF_L/DF_{NW}>n^*$, respectively. Weak waves correspond to $|S_{local}|>1$ or $RHS/LHS<1$ and are marked by subindex $L$, whereas intense waves correspond to $|S_{local}|<1$ and  $RHS/LHS>1$ and are marked by subindex $NL$.}
\end{figure}

For each wave measurement, we obtained the concurrently measured electron distribution function $DF(E,\alpha)$. We separated these DF measurements into three categories according to the wave properties: DFs associated with linear ($DF_L$) wave particle interactions, DFs associated with nonlinear ($DF_{NL}$) interactions, and DFs without significant whistler-mode waves observed ($DF_{NW}$). All electron distribution functions were normalized to the local plasma density to suppress any effects related to the strong density variations across the foreshock transients. In each transient event, we first calculated the median values of  $DF_{NL}$, $DF_{L}$, and $DF_{NW}$, and their ratios ($DF_{L}/DF_{NW}$ and $DF_{NL}/DF_{NW}$). Then we calculated the median values of these ratios. The results  are shown in Figure \ref{fig5}(a,b); they are plotted against energy (normalized to $E_R$) and pitch angle. There is a clear phase space density increase around and above the resonance energy and $\alpha\sim90^\circ$ for $DF_L$. This increase may be due to a combination of an initial electron anisotropy driving whistler-mode wave generation and electron acceleration by waves. The increase in phase space density is much more localized in energy (around $E_R$) and covers a wider $\alpha$ range for $DF_{NL}$. If the strong $DF_{NL}$ peak around $\alpha\sim 90^\circ$ is due to the strong initial anisotropy needed for intense wave generation, the $DF_{NL}$ increase at small pitch-angles ($\alpha<45^\circ$ and $\alpha>135^\circ$) is most likely due to the effective electron mixing by nonlinear resonances with waves \citep[see discussion of nonlinear resonant effects in, e.g.,][]{Vainchtein18:jgr}. A weak decrease of $DF_{L}/DF_{NW}$ and $DF_{NL}/DF_{NW}$ at energies well below the resonance energy is unlikely to be related to wave-particle resonant interactions but could be due to preferential wave generation within hot plasma regions where the cold electron density is reduced. The black lines in Figure \ref{fig5}(a,b) show the contours of $DF_{NW}$. The results with $DF_{NW}<10^{-5}$ are not statistically significant because such a small phase space density may lead to large errors. 

To further investigate the difference between the distributions of phase space density associated with weak and intense waves, we plot the probability distributions of $DF_L/DF_{NW}>n^*$ (for weak waves) and $DF_{NL}/DF_{NW}>n^*$ (for intense waves) over all pitch angles, where $n^*$ stands for the value of the ratio. Figures \ref{fig5}(c,d) show the percentage of events with  $DF_L/DF_{NW}>n^*$ and $DF_{NL}/DF_{NW}>n^*$ at different energies, respectively. There is a clear difference between weak and intense waves. For $DF_L/DF_{NW}>n^*$ around the resonance energy the probability distribution is reduced significantly for $n^*>2$, i.e. there is nearly negligible probability of observing a phase space density increase by a factor of $>2$ in association with weak waves. For $DF_{NL}/DF_{NW}>n^*$ around the resonance energy the probability distribution remains large even for $n^*\sim 3$, i.e., there is a significant probability of observing intense waves in association with resonant phase space density increase by a factor of $\sim 3$. Moreover, Figure \ref{fig5}(d) shows that the probability distribution has a local maximum around the resonance energy.  These results are consistent with our assertion of a significant contribution of nonlinear resonant interactions to electron acceleration.

\section{Summary and Discussion}\label{sec:conclusions}
We statistically studied whistler wave properties in foreshock transients, using a database of 208 transient events comprised of HFAs or FBs.  We also investigated the regimes of wave particle interactions and the effects of nonlinear interactions on electron distributions. Specifically, we showed that:

\begin{enumerate}
        \item [1.] Whistler waves exist in $85$\% of the foreshock transients examined. These waves are most often seen around the edge of the core or the compressional boundary regions of foreshock transients. 
        \item [2.] The whistler waves in foreshock transients have frequencies around $0.2f_{ce}$, regardless of the abrupt change of the background magnetic field at their location. This indicates that they are generated locally, i.e., they do not propagate to the satellite from a distinctly different location in the transient or its vicinity. On average, the whistler waves are quasi-parallel. However, the waves in the core region tend to be more oblique than the waves in the compressional boundary. 
        \item [3.] Intense whistler waves are frequently observed. Their median amplitude is around $0.01B_0$ ($\sim10s-100s$ pT). The resonance energy for electrons is around $10s-100s$ eV, and $15$\% of the observed whistler waves are sufficiently intense and narrow-band to resonate with electrons nonlinearly. 
        \item [4.] Events with potential nonlinear wave-particle interactions show a clear increase in phase space density around the resonance energy. This increase is larger than that for observations associated with low-intensity waves. This suggests that nonlinear resonant interactions can contribute significantly to electron acceleration.
\end{enumerate}

Recent observations of energetic electrons in foreshock transients \citep[e.g.,][]{wilson2016_foreshock, Liu2017} are highly suggestive that such transients can accelerate electrons and may provide a seed population for further acceleration at the bow shock - the main source of energetic particles at the dayside. Foreshock transients also exist at other high Mach number quasi-parallel shocks (e.g., at outer planets, where the solar wind Mach number is high, or occasionally even at Mars, \cite{Collionson2015GeoRL.}). The whistler waves studied are therefore, by analogy, likely common and significant for electron acceleration inside foreshock transients in planets and in other astrophysical contexts (such as at supernova shocks which can produce cosmic rays). Our statistical results on the whistler-mode wave intensity demonstrate that a significant portion of the observed waves resonates with electrons nonlinearly. The nonlinear interactions can play an important role in electron acceleration processes in the following way: 

First, nonlinear interactions have diffusion rates that are different from (often faster than) those found in classical, linear theory. We have shown that intense waves can significantly modify electron distributions, which indicates that they can alter electron trajectories. This invalidates the approximation of unperturbed trajectories for classical scaling of electron pitch-angle diffusion rate $D\propto \langle B_w\rangle ^2$ \citep{Kennel&Engelmann66}, and therefore, the classical pitch-angle diffusion rate does not apply. If the waves propagate in short wave packets (containing only a few wave periods each; see, e.g., such wave packets in the bow shock in \citet{Hull12,Oka17,Oka19}), then the main nonlinear effect will be the change of the diffusion rate scaling, $D\propto\langle B_w\rangle^{1/2}$ \citep[e.g.,][]{Artemyev21:pre}. Thus, a simple extrapolation of quasi-linear theory scaling, $D\propto \langle B_w\rangle ^2$, to high-intensity waves would significantly overestimate the actual diffusion rate. The diffusion rate is an important element of the stochastic shock drift acceleration model \citep{Amano20}, and therefore the change in the scaling of $D$  may modify the efficiency of the resultant electron acceleration in this model.

Second, intense waves propagating in long wave packets (each containing tens of wave periods) may result in nonlinear resonant acceleration of electrons in an inhomogeneous magnetic field via phase trapping \citep[see reviews by][]{Shklyar09:review,Albert13:AGU}. To be effective, this acceleration mechanism should be combined with electron periodic motions in magnetic field traps, i.e., electrons should experience multiple resonant interactions. The general magnetic field configuration of foreshock transients does allow such trapping \citep{Liu19:foreshock}. Moreover, such trapping can also be provided by ultra-low-frequency compressional magnetic field fluctuations \citep[e.g.,][]{Oka19, Lichko&Egedal20} and by the transient-bow shock magnetic field configuration \citep{Liu17:foreshock&electrons,Turner2018}. Therefore, foreshock transients embedding intense whistler-mode waves (resonating with electrons nonlinearly) may serve as effective electron accelerators, if the interplanetary magnetic field can trap electrons. Further theoretical analysis of nonlinear wave-particle resonances and observational analysis of energetic electron bursts and whistler waves associated with foreshock transients may reveal the efficiency of such nonlinear resonant acceleration.

\section*{Acknowledgements}
A.A. acknowledges the NASA HGI project 80NSSC21K0581. X.-F. S. and T. Z. L. acknowledge NSF award AGS-1941012. 
MMS data were downloaded from https://lasp.colorado.edu/mms/sdc/public. Data access and processing was done using SPEDAS V4.1, see \citet{Angelopoulos19}.

\bibliography{full,addon}

\end{document}